\documentclass[12pt]{article}
\usepackage{amsmath,amsfonts,amssymb,graphicx}
\usepackage{graphics, setspace}
\usepackage{cite} 
\usepackage{hyperref}
\usepackage{epstopdf}
\DeclareGraphicsExtensions{.eps}

\begin{document}
\thispagestyle{empty}

\def\theequation{\arabic{section}.\arabic{equation}}
\def\a{\alpha}
\def\b{\beta}
\def\g{\gamma}
\def\d{\delta}
\def\dd{\rm d}
\def\e{\epsilon}
\def\ve{\varepsilon}
\def\z{\zeta}
\def\B{\mbox{\bf B}}\def\cp{\mathbb {CP}^3}

\newcommand{\h}{\hspace{0.5cm}}

\begin{titlepage}

\renewcommand{\thefootnote}{\fnsymbol{footnote}}
\begin{center}
{\Large \bf Current algebra for a generalized two-site Bose-Hubbard model}
\end{center}
\vskip 1.2cm \centerline{\bf Gilberto N. Santos Filho$^{1}$}

\vskip 10mm
\centerline{\sl$\ ^1$ Centro Brasileiro de Pesquisas F\'{\i}sicas - CBPF}
\centerline{\sl Rua Dr. Xavier Sigaud, 150, Urca, Rio de Janeiro - RJ - Brazil}
\vskip .5cm

\centerline{\tt gfilho@cbpf.br}
 
\vskip 20mm

\baselineskip 18pt

\begin{center}
{\bf Abstract}
\end{center}

I present a current algebra for a generalized two-site Bose-Hubbard model and use it to get the quantum dynamics of the currents. For different choices of the Hamiltonian parameters we get different currents dynamics. I  generalize  the Heisenberg equation of motion to write the $n$-th time derivative of any operator.

\end{titlepage}
\newpage
\baselineskip 18pt

\def\nn{\nonumber}
\def\tr{{\rm tr}\,}
\def\p{\partial}
\newcommand{\non}{\nonumber}
\newcommand{\bea}{\begin{eqnarray}}
\newcommand{\eea}{\end{eqnarray}}
\newcommand{\bde}{{\bf e}}
\renewcommand{\thefootnote}{\fnsymbol{footnote}}
\newcommand{\be}{\begin{eqnarray}}
\newcommand{\ee}{\end{eqnarray}}

\vskip 0cm

\renewcommand{\thefootnote}{\arabic{footnote}}
\setcounter{footnote}{0}

\setcounter{equation}{0}
\section{Introduction}

Since the first experimental verification of the Bose-Einstein condensation (BEC) \cite{ak,anderson,wwcch} occurred more then seven decades after its theoretical prediction \cite{bose,eins} a great deal of progress has been made in the theoretical and experimental study of this many body physical phenomenon \cite{Hulet,Dalfovo,l01,Bagnato,cw,Donley,Piza,Bloch1,Bloch2,Bloch3,Bloch4,Minardi,Greiner,Greene,Kraemer,
Oberthaler4,Riedel,Kinoshita,Thierry,Inouye,Carusotto}. Looking in this direction a laser was used in an experiment to divide a BEC in two parts to study the interference phenomenon between two BECs \cite{Ketterle1,Ketterle2}. These two BECs can be coupled by Josephson tunnelling \cite{Oberthaler1,Oberthaler2,Oberthaler3,Levy,Bloch5,Josephson1,Josephson2} with the atoms tunnelling between the condensates in the same way that a Cooper's pair in a superconductor Josephson junction. This system is equivalent to a two-wells system with the particles tunnelling across a barrier between the wells. To study this system a model, known as the {\it canonical Josephson Hamiltonian}, was proposed by Leggett \cite{l01}. Since then many models have been used to study the BECs such as the quantum dynamics of the tunnelling of atoms between the two condensates,  the entanglement, the quantum phase transitions, the classical analysis, the atom-molecule interconversion and the quantum metrology \cite{Javanainen,l02,milb,hines2,ours,our,hines,sarma,GSantos06a,GSantos09,GSantos10,Gross}. The algebraic Bethe ansatz method has been used to solve and study some of these models \cite{GSantos11a,GSantos11b,GSantos13,jlletter,jlreview,jlsigma,Angela,ATonel,GSantos06b,Angela2,GSantosBVS-arxiv,jldensity}.  I will consider here a generalized issue of the models \cite{l01,jlreview} by the introducing of the on-well energies and leaving free choice for the interaction parameters that also permits the study of the tunnelling between two condensates  with atoms of different species (different chemical elements) or atoms in different states in each condensate. The on-well energies is  determined by the internal states of the atoms in the condensates, by the kinetic and interaction energies of the atoms and/or the external potentials. I will study in this work the current algebra and  the quantum dynamics of the currents for this model using a generalization of the Heisenberg equation of motion that make possible to write the second time derivative of the current operators. The generalized model is described by the Hamiltonian 
\begin{eqnarray}
\hat{H} &=& \sum_{i,j=1}^2 K_{ij}\hat{N}_i\hat{N}_j  - \sum_{i=1}^2 (U_i - \mu_i)\hat{N}_i  - \sum_{i,j=1\atop i\neq j}^2 \Omega_{ij} \hat{a}_i^\dagger \hat{a}_j,
\label{ham} 
\end{eqnarray}
\noindent where, $\hat{a}_i^\dagger (\hat{a}_i)$, denote the single-particle creation (annihilation) operators  and $\hat{N}_i = \hat{a}_i^\dagger \hat{a}_i$ are the corresponding  boson number operators in each condensate. The boson operator total number of particles, $\hat{N} = \hat{N}_1+\hat{N}_2$,  is a conserved quantity, $[\hat{H},\hat{N}]=0$. The couplings $K_{ij}$, with $K_{ij} = K_{ji} \; (i \neq j)$, provides the interaction strength between the  bosons and they are proportional to the $s$-wave scattering length,  $\Omega_{ij}$ are the   amplitude of tunnelling,  $\mu_i$ are the external potentials and $U_i = K_{ii} - \kappa_i$ are the on-well energies per particle,  with $\kappa_i$ the kinetic energies in each condensate.

For the particular choice of the couplings parameters we can get some Hamiltonians, as for example  by the choices  $K_{ii} = \kappa_i = \frac{K}{8}$, $ K_{12} = -\frac{K}{8}$, $\Delta\mu = \mu_1 - \mu_2 = 2\mu$ and $\Omega_{12} = \Omega_{21} = \frac{{\cal E}_{J}}{2}$ we get the canonical Josephson Hamiltonian studied in  \cite{l01}.  The  case with  $K_{12} = \kappa_i = 0$, $U_i = K_{ii} = U/2$, $\epsilon = \mu_1 - \mu_2 = 2\mu$, and $\Omega_{12} = \Omega_{21} = t$ was used  to study the interplay between disorder and interaction \cite{sarma}.   We can control the tunnelling using the external potentials and we have a symmetric two-wells if  $\Delta\mu = 0$ and when we turn on $\Delta\mu$ we break the symmetry. For the symmetric case we also can put $\mu_1=\mu_2=\mu$ and change the deep of both wells at the same time. This mean that we also can adjust the on-well energies using the external potential in the symmetric case. In the antisymmetric case $\Delta\mu\neq0$ we can change the bias of one well and increase the on-well energy. In this case it is called a tilted two-wells potential \cite{Dounas,ours}.  

The paper is organized as follows. In section 2, I  discuss briefly the symmetries of the model. In section 3, I present the current algebra. In section 4, I present a generalization of the Heisenberg equation of motion, get the quantum dynamics of the currents and compare with the experiments. In section 5, I summarize the results.

\section{Symmetries}

 The Hamiltonian (\ref{ham}) is  invariant under the $\mathbb{Z}_2$ mirror transformation $\hat{a}_j \rightarrow -\hat{a}_j, \hat{a}_j^{\dagger} \rightarrow -\hat{a}_j^{\dagger}$, and under the global $U(1)$  gauge transformation $\hat{a}_j \rightarrow e^{i\alpha}\hat{a}_j$, where $\alpha$ is an arbitrary $c$-number and $\hat{a}^{\dagger}_j\rightarrow e^{-i\alpha}\hat{a}^{\dagger}_j,\;\;j=1,2$. For $\alpha = \pi$ we get again the $\mathbb{Z}_2$ symmetry. The global $U(1)$  gauge invariance is associated with the conservation of the total number of atoms $\hat{N} = \hat{N}_1 + \hat{N}_2$ and the $\mathbb{Z}_2$ symmetry is associated with the parity of the wave function by the relation

\begin{equation}
 \hat{P} \; |\Psi\rangle = (-1)^N |\Psi\rangle,
\end{equation}  

\begin{equation}
|\Psi\rangle = \sum_{n = 0}^N \; C_{n,N-n} \frac{(\hat{a}_1^{\dagger})^{n}}{\sqrt{n!}} \frac{(\hat{a}_2^{\dagger})^{N - n}}{\sqrt{(N - n)!}} |0,0\rangle, \label{wf1}
\end{equation} 
\noindent where $\hat{P}$ is the parity operator and $[\hat{H},\hat{P}]=0$.

\begin{figure}[hb]
\begin{center}
\includegraphics[scale=0.2]{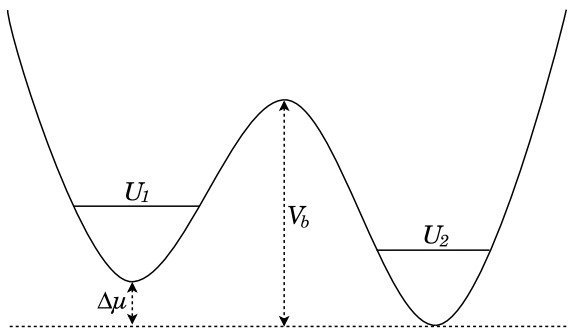}
\caption{Two-well potential showing the tunnelling for $U_1 = U_2$ and $\Delta\mu \neq 0$ with the height of the barrier $V_b$. }
\label{tw}
\end{center}
\end{figure}

 There is also the permutation symmetry of the atoms of the two wells if we have $\Delta\mu = 0$, $U_1 =  U_2$ and $\Omega_{12} = \Omega_{21}$. When we turn on $\Delta\mu$ or put $U_1 \neq  U_2$ or $\Omega_{12} \neq \Omega_{21}$ we break the symmetry. The wave function (\ref{wf1}) is symmetric under this permutation
 \begin{equation}
 \hat{{\cal P}}\; |\Psi\rangle = \sum_{n = 0}^N \;  C_{N - n,n} \frac{(\hat{a}_1^{\dagger})^{N - n}}{\sqrt{(N - n)!}}\frac{(\hat{a}_2^{\dagger})^{n}}{\sqrt{n!}} |0,0\rangle = |\Psi\rangle,
\end{equation}  
\noindent where $\hat{{\cal P}}$ is the permutation operator and $[\hat{H},\hat{{\cal P}}]=0$ if $\Delta\mu = 0$ \cite{jlletter}, $U_1 =  U_2$ and $\Omega_{12} = \Omega_{21}$. In the Fig. \ref{tw} we represent the two BEC by a two-well potential for the case  $\Delta\mu \neq 0$ and $U_1 =  U_2$.

 The symmetries of the Hamiltonian (\ref{ham}) imply  degeneracy. For the conservancy of $\hat{N}$ we have  that all wave function of the Hamiltonian (\ref{ham})  are degenerated eigenfunctions of $\hat{N}$ with the same eigenvalue $N$.  For the parity operator $\hat{P}$ all wave function of the Hamiltonian (\ref{ham})  are even or odd depending if $N$ is even or odd. All wave functions are degenerated eigenfunctions of $\hat{P}$ with the same eigenvalue $\lambda = +1$ if $N$ is even  or they are degenerated eigenfunctions of $\hat{P}$ with the same eigenvalue $\lambda = -1$ if $N$ is odd. For the permutation operator $\hat{{\cal P}}$ all wave function of the Hamiltonian (\ref{ham}) are degenerated eigenfunctions with the same eigenvalue $\lambda = +1$. 

\section{Current Algebra}

The quantum dynamics of any operator $\hat{O}$ in the Heisenberg picture is determined by the Heisenberg equation of motion
\begin{equation}
 \frac{d\hat{O}}{dt} = \frac{i}{\hbar}[\hat{H},\hat{O}].
 \label{dOdtH}
\end{equation}

The boson operator total number of particles, $\hat{N} = \hat{N}_1+\hat{N}_2$, 
is a conserved quantity, $[\hat{H},\hat{N}]=0$, and it is commutable compatible operator (CCO) with the boson operators number of particles in each well, $[\hat{N},\hat{N}_1]=[\hat{N},\hat{N}_2]= [\hat{N}_1,\hat{N}_2]=0$. The boson operators number of particles in each well don't commute with the Hamiltonian and their time evolution is determined by the Josephson tunnelling current operator,

\begin{equation}
\hat{\mathcal{J}} = \frac{1}{2i} (\hat{a}_1^\dagger \hat{a}_2 - \hat{a}_2^\dagger \hat{a}_1), \label{cj}
\end{equation}
\noindent in coherent opposite phases because of the conservancy of $\hat{N}$, with 
\begin{equation}
[\hat{H},\hat{N}_1]= +2i\Omega\hat{\mathcal{J}}, ~~~ [\hat{H},\hat{N}_2]= -2i\Omega\hat{\mathcal{J}},
\label{hn}
\end{equation}
\noindent and 
\begin{equation}
\frac{d\hat{N}_1}{dt}  =  -2\frac{\Omega}{\hbar}\hat{\mathcal{J}}, \label{CDN1}
\end{equation}

\begin{equation}
\frac{d\hat{N}_2}{dt}  =  + 2\frac{\Omega}{\hbar}\hat{\mathcal{J}}. \label{CDN2}
\end{equation}
\noindent Hereafter and in the Eqs. (\ref{hn}), (\ref{CDN1}) and (\ref{CDN2})  above,  we will consider $\Omega_{12} = \Omega_{21}=\Omega$.

If we introduce a phase $\phi_{ij}$ for each term $\hat{a}_i^\dagger \hat{a}_j$, $i,j = 1,2$, we can write the current (\ref{cj}) as 
 
\begin{equation}
\hat{\mathcal{J}} = \frac{1}{2} (e^{i\phi_{12}}\hat{a}_1^\dagger \hat{a}_2 + e^{i\phi_{21}}\hat{a}_2^\dagger \hat{a}_1),\label{cj1}
\end{equation}
\noindent with $\phi_{12}=3\pi/2$ and $\phi_{21}=\pi/2$. So, the phase difference in the current $\hat{\mathcal{J}}$ is $|\Delta\phi| = \pi$. 
   The tunnelling current $\hat{\mathcal{J}}$ together with the imbalance current $\hat{\mathcal{I}}$
\begin{equation}
\hat{\mathcal{I}} = \frac{1}{2}(e^{i\phi_{11}}\hat{N}_1 + e^{i\phi_{22}}\hat{N}_2),\label{ci1}
\end{equation}
\noindent with $\phi_{11}=0$ and $\phi_{22}=\pi$, to get the phase difference in the current $\hat{\mathcal{I}}$ equal to $|\Delta\phi| = \pi$, and the coherent correlation tunnelling current operator $\hat{\mathcal{T}}$
\begin{equation}
\hat{\mathcal{T}} = \frac{1}{2}(e^{i\phi_{12}}\hat{a}_1^\dagger \hat{a}_2 + e^{i\phi_{21}}\hat{a}_2^\dagger \hat{a}_1),\label{ct1}
\end{equation}
\noindent with $\phi_{12} = 0 \;\mbox{or}\; 2\pi$ and $\phi_{21} = 0 \;\mbox{or}\; 2\pi$, to get the phase difference in the current $\hat{\mathcal{T}}$ equal to $|\Delta\phi| = 0 \;\mbox{or}\; 2\pi$, generates the currents algebra
\begin{equation}
 [\hat{\mathcal{T}},\hat{\mathcal{J}}]= +i\hat{\mathcal{I}}, ~~~ [\hat{\mathcal{T}},\hat{\mathcal{I}}]= -i\hat{\mathcal{J}}, ~~~ [\hat{\mathcal{J}},\hat{\mathcal{I}}]= +i\hat{\mathcal{T}}.
\label{CA1} 
\end{equation}
\noindent With the identification $\hat{L}_x \equiv  \hbar\hat{\mathcal{T}}$, $\hat{L}_y \equiv  \hbar\hat{\mathcal{J}}$, and $\hat{L}_z \equiv \hbar\hat{\mathcal{I}}$ we can write it in the standard compact way of the momentum angular
\begin{equation}
 [\hat{L}_k,\hat{L}_l] = i \hbar \varepsilon_{klm}\hat{L}_m,
\end{equation}
\noindent where $\varepsilon_{klm}$ is the  antisymmetric Levi-Civita tensor with $k,l,m=x,y,z$ and $\varepsilon_{xyz} = +1$.

 We have two Casimir operators for that currents algebra. One of them is the total number of particles, $\hat{\mathcal{\mathfrak{C}}}_{1} = \hat{N}$, related to the $U(1)$ symmetry and the another one is related to the momentum angular algebra and the $O(3)$ symmetry, $\hat{\mathcal{\mathfrak{C}}}_{2} = \hat{\mathcal{T}}^2 + \hat{\mathcal{I}}^2 + \hat{\mathcal{J}}^2$.

 We can show that $\hat{\mathcal{\mathfrak{C}}}_{2}$ is just a function of $\hat{\mathcal{\mathfrak{C}}}_{1}$
\begin{equation}
\hat{\mathcal{\mathfrak{C}}}_{2} = \frac{\hat{\mathcal{\mathfrak{C}}}_{1}}{2}\left(\frac{\hat{\mathcal{\mathfrak{C}}}_{1}}{2} +  1\right).
\end{equation}


The Casimir operators $\hat{\mathcal{\mathfrak{C}}}_{1}$ and $\hat{\mathcal{\mathfrak{C}}}_{2}$, the boson number of particles in each well $\hat{N}_1$, $\hat{N}_2$,  and the imbalance current operator, $\hat{\mathcal{I}}$, are CCO and so they have the same set of eigenfunctions and can simultaneous have well defined  values

\begin{equation}
\hat{\mathcal{\mathfrak{C}}}_{2} |n_1,n_2\rangle = \frac{N}{2}\left(\frac{N}{2} +  1\right) |n_1,n_2\rangle,
\end{equation}

\begin{equation}
\hat{\mathcal{I}} |n_1,n_2\rangle = \frac{1}{2} \left(n_1 - n_2\right) |n_1,n_2\rangle.
\end{equation}

We also can use the realization of the $SU(2)$ algebra

\begin{equation}
\hat{\mathcal{L}}_{\pm} = \frac{1}{\hbar}(\hat{L}_{x} \pm i\hat{L}_{y}), ~~~~~ \hat{\mathcal{L}}_{z} = \frac{1}{\hbar}\hat{L}_{z},
\end{equation}

\noindent with the commutation relations
\begin{equation}
[\hat{\mathcal{L}}_{z}, \hat{\mathcal{L}}_{\pm}]  =  \pm \hat{\mathcal{L}}_{\pm}, ~~~~~  
[\hat{\mathcal{L}}_{+}, \hat{\mathcal{L}}_{-}]  =   2\hat{\mathcal{L}}_{z},
\end{equation}
\noindent that we can write as

\begin{equation}
[\hat{\mathcal{L}}_{k},\hat{\mathcal{L}}_{l}] = \varepsilon_{kl-}\hat{\mathcal{L}}_{+} + \varepsilon_{kl+}\hat{\mathcal{L}}_{-} + 2 \varepsilon_{zkl}\hat{\mathcal{L}}_{z},
\end{equation}
\noindent  with $k,l=z,+,-$ and $\varepsilon_{z+-} = +1$.

The $SU(2)$ algebra has three Casimr operators, $\hat{\mathcal{\mathfrak{C}}}_{1}$ and 

\begin{equation}
\hat{\mathcal{\mathfrak{C}}}_{3} = \hat{\mathcal{L}}_{+}\hat{\mathcal{L}}_{-} + \hat{\mathcal{L}}_{z}^{2} - \hat{\mathcal{L}}_{z},
\end{equation}

\begin{equation}
\hat{\mathcal{\mathfrak{C}}}_{4} = \hat{\mathcal{L}}_{-}\hat{\mathcal{L}}_{+} + \hat{\mathcal{L}}_{z}^{2} + \hat{\mathcal{L}}_{z}.
\end{equation}

We can show that these Casimir operators are equals to $\hat{\mathcal{\mathfrak{C}}}_{2}$. In the deformed $SU(2)$ and $O(3)$ algebras they are different \cite{Graefe}.



Using the commutation relations of the currents  (\ref{CA1}) it is easy to calculate the anticommutators

\begin{equation}
[\hat{\mathcal{T}},\hat{\mathcal{I}}]_{+}  =  2 \hat{\mathcal{I}} \hat{\mathcal{T}} - i\hat{\mathcal{J}},  \label{AC1a}
\end{equation}

\begin{equation}
[\hat{\mathcal{T}},\hat{\mathcal{J}}]_{+}  =  2 \hat{\mathcal{J}} \hat{\mathcal{T}} + i   \hat{\mathcal{I}}, \label{AC2a}
\end{equation}

\begin{equation}
[\hat{\mathcal{J}},\hat{\mathcal{I}}]_{+}  =  2 \hat{\mathcal{I}} \hat{\mathcal{J}} + i\hat{\mathcal{T}}.\label{AC3a}
\end{equation} 
\noindent We will use these anticommutators together with the commutators (\ref{CA1}) in the calculus of the currents quantum dynamics.  The current algebra (\ref{CA1}) is the same for the model \cite{l01} that was   described  in \cite{GSantosCA1}.

\section{Current Quantum Dynamics} 

 We can rewrite the Hamiltonian (\ref{ham}) using those currents operators
\begin{eqnarray}
\hat{H} & = & \alpha\hat{\mathcal{I}}^2 + \hat{\mathcal{\mathfrak{Z}}}\hat{\mathcal{I}} 
 - 2\Omega\hat{\mathcal{T}} + \frac{\hat{\mathcal{\mathfrak{C}}}_{1}}{2}\left(\frac{\hat{\mathcal{\mathfrak{C}}}_{1}}{2}\rho +  \xi\right),
\label{ham2} 
\end{eqnarray}
\noindent where
\begin{eqnarray}
\alpha & = & K_{11} - 2K_{12} + K_{22}, \nonumber \\
\beta  & = & K_{11} - K_{22}, \nonumber \\
\gamma & = & \mu_1 - U_1 - \mu_2 + U_2, \nonumber \\ 
\rho & = & K_{11} + 2K_{12} + K_{22}, \nonumber \\
\xi   & = & \mu_1 - U_1 + \mu_2 - U_2. 
\end{eqnarray}

\noindent We have defined the Casimir operator $\hat{\mathcal{\mathfrak{Z}}} = \beta\hat{\mathcal{\mathfrak{C}}}_{1} + \gamma$ and we can see that the Casimir operators are also  conserved quantities, $[\hat{H},\hat{\mathcal{\mathfrak{C}}}_{1}]=0$.



The quantum dynamics of the currents (\ref{cj1}), (\ref{ci1}) and (\ref{ct1}) is determined by the current  algebra, their commutation relations with the Hamiltonian and the parameters. We can use the Heisenberg equation of motion (\ref{dOdtH}) to write the second time derivative of any operator $\hat{O}$ in the Heisenberg picture as \cite{GSantosCA1}
\begin{equation}
 \frac{d^2\hat{O}}{dt^2} = \left(\frac{i}{\hbar}\right)^2 [\hat{H},[\hat{H},\hat{O}]],
 \label{dOdt0}
\end{equation}
\noindent or as
\begin{equation}
 \frac{d^2\hat{O}}{dt^2} = \frac{i}{\hbar}[\hat{H},\frac{d\hat{O}}{dt}].
 \label{dOdt}
\end{equation}

\noindent It is direct to generalize the Eqs. (\ref{dOdt0}) and (\ref{dOdt}) for the $n$-th  time derivative of any operator $\hat{O}$ in the Heisenberg picture. So we can write

\begin{equation}
 \frac{d^n\hat{O}}{dt^n} = \left(\frac{i}{\hbar}\right)^n \underbrace{[\hat{H},[\hat{H},[\hat{H},\ldots,[\hat{H},\hat{O}]]]}_{n\;commutators},
 \label{dndtnc}
\end{equation}
\noindent or as
\begin{equation}
 \frac{d^n\hat{O}}{dt^n} = \frac{i}{\hbar}[\hat{H},\frac{d^{n-1}\hat{O}}{dt^{n-1}}],
 \label{dndtn}
\end{equation}
\noindent where we have defined

\begin{equation}
 \frac{d^0\hat{O}}{dt^0} \equiv \hat{O},
 \label{dOdtd}
\end{equation}
\noindent and $ n\geq 1$. We get the Heisenberg equation of motion (\ref{dOdtH}) for $n=1$ and the Eqs. (\ref{dOdt0}) and (\ref{dOdt}) for $n=2$. Using the Eq. (\ref{dOdt0}) or (\ref{dOdt}) we found the following equations for the quantum dynamics of the currents

\begin{eqnarray}
\frac{d^{2}\hat{\mathcal{I}}}{dt^{2}}+4\frac{\Omega^{2}}{\hbar^{2}}\hat{\mathcal{I}} & = & -4\frac{\Omega\alpha}{\hbar^{2}}\hat{\mathcal{I}}\hat{\mathcal{T}}+2i\frac{\Omega\alpha}{\hbar^{2}}\hat{\mathcal{J}}-2\frac{\Omega}{\hbar^{2}}\hat{\mathcal{\mathfrak{Z}}}\hat{\mathcal{T}},\label{eq1:wideeq}
\end{eqnarray}

\begin{eqnarray}
\frac{d^{2}\hat{\mathcal{J}}}{dt^{2}}+\frac{1}{\hbar^{2}}\left[\alpha^{2}+\hat{\mathcal{\mathfrak{Z}}}^{2}+4\Omega^{2}\right]\hat{\mathcal{J}} & = & -4\frac{\alpha^{2}}{\hbar^{2}}\hat{\mathcal{I}}^{2}\hat{\mathcal{J}}-2i\frac{\alpha^{2}}{\hbar^{2}}\hat{\mathcal{I}}\hat{\mathcal{T}} 
- 2\frac{\alpha}{\hbar^{2}}\hat{\mathcal{\mathfrak{Z}}}\hat{\mathcal{I}}\hat{\mathcal{J}} \nonumber\\ & - & 4\frac{\alpha\Omega}{\hbar^{2}}\hat{\mathcal{J}}\hat{\mathcal{T}}  -  2i\frac{\alpha}{\hbar^{2}}\hat{\mathcal{\mathfrak{Z}}}\hat{\mathcal{T}}-2i\frac{\alpha\Omega}{\hbar^{2}}\hat{\mathcal{I}}, \label{eq2:wideeq}
\end{eqnarray}

\begin{eqnarray}
\frac{d^{2}\hat{\mathcal{T}}}{dt^{2}}+\frac{1}{\hbar^{2}}\left(\alpha^{2}+\hat{\mathcal{\mathfrak{Z}}}^{2}\right)\hat{\mathcal{T}} & = & -4\frac{\alpha^{2}}{\hbar^{2}}\hat{\mathcal{I}}\hat{\mathcal{I}}\hat{\mathcal{T}}+4i\frac{\alpha^{2}}{\hbar^{2}}\hat{\mathcal{I}}\hat{\mathcal{J}}-4\frac{\alpha}{\hbar^{2}}\hat{\mathcal{\mathfrak{Z}}}\hat{\mathcal{I}}\hat{\mathcal{T}} \nonumber\\
 & + & 2i\frac{\alpha}{\hbar^{2}}\hat{\mathcal{\mathfrak{Z}}}\hat{\mathcal{J}}-4\frac{\Omega\alpha}{\hbar^{2}}(\hat{\mathcal{I}}^{2}-\hat{\mathcal{J}}^{2})-2\frac{\Omega}{\hbar^{2}}\hat{\mathcal{\mathfrak{Z}}}\hat{\mathcal{I}}.
\label{eq3:wideeq}
\end{eqnarray}

We can see from the Eqs. (\ref{eq1:wideeq}), (\ref{eq2:wideeq}) and (\ref{eq3:wideeq}) that  the currents are coupled on the right hand side of these equations. To simplify our analysis we will make some choices of the parameters. Different choices of the  parameters of the Hamiltonian gives us different dynamics for the currents. Fortunately the  parameters appear in these equations in the linear and quadratic power. So we can consider a perturbation theory in the parameters of the Hamiltonian until the second power  terms. If we calculate the $n$-th time derivative of the current operators we will get the $n$-th power of  the  parameters. Here we will need consider only until second order time derivative of the current operators. We can try to use mean field theory (MFT) to decouple the currents to get some insight. In the first approximation, for example, we can use $\langle \hat{L}_k\hat{L}_l\rangle \approx \langle \hat{L}_k\rangle\langle\hat{L}_l\rangle$. But for this approximation we get from the commutation relations (\ref{CA1}) that $\langle \hat{\mathcal{I}}\rangle \approx \langle \hat{\mathcal{J}}\rangle \approx \langle \hat{\mathcal{T}}\rangle \approx 0$.  Therefore, the currents are correlated by the currents algebra (\ref{CA1}) that forbid MFT even in the first approximation. We also can see the  correlation  between the currents, using the currents algebra (\ref{CA1}), writing the  Heisenberg uncertainty relations for each couple of currents

\begin{eqnarray}
\langle (\widehat{\Delta \mathcal{T}})^2 \rangle\langle (\widehat{\Delta \mathcal{J}})^2\rangle & \leq & \frac{1}{4} \langle \hat{\mathcal{I}} \rangle^2, \\
\langle (\widehat{\Delta \mathcal{T}})^2 \rangle\langle (\widehat{\Delta \mathcal{I}})^2\rangle & \leq & \frac{1}{4} \langle \hat{\mathcal{J}} \rangle^2, \\
\langle (\widehat{\Delta \mathcal{J}})^2 \rangle\langle (\widehat{\Delta \mathcal{I}})^2\rangle & \leq & \frac{1}{4} \langle \hat{\mathcal{T}} \rangle^2,
\label{HUR}
\end{eqnarray}
\noindent where we are introducing the  operator $\widehat{\Delta L}_k = \hat{L}_k - \langle \hat{L}_k \rangle$.

Choosing $\alpha=\beta=0$ we get  three linear second order differential equations 

\begin{eqnarray}
\frac{d^{2}\hat{\mathcal{I}}}{dt^{2}}+4\frac{\Omega^{2}}{\hbar^{2}}\hat{\mathcal{I}} & = & -2\frac{\Omega\gamma}{\hbar^{2}}\hat{\mathcal{T}},
\label{eq1:wideeq-alf} 
\end{eqnarray}

\begin{eqnarray}
\frac{d^{2}\hat{\mathcal{J}}}{dt^{2}} + \frac{1}{\hbar^{2}}\left(\gamma^{2} + 4\Omega^{2}\right)\hat{\mathcal{J}} & = & 0,
\label{eq2:wideeq-alf} 
\end{eqnarray}

\begin{eqnarray}
\frac{d^{2}\hat{\mathcal{T}}}{dt^{2}} + \frac{\gamma^{2}}{\hbar^{2}}\hat{\mathcal{T}} & = & -2\frac{\Omega\gamma}{\hbar^{2}}\hat{\mathcal{I}}.
\label{eq3:wideeq-alf}
\end{eqnarray}

\begin{figure}[h]
\begin{center}
\includegraphics[scale=0.7]{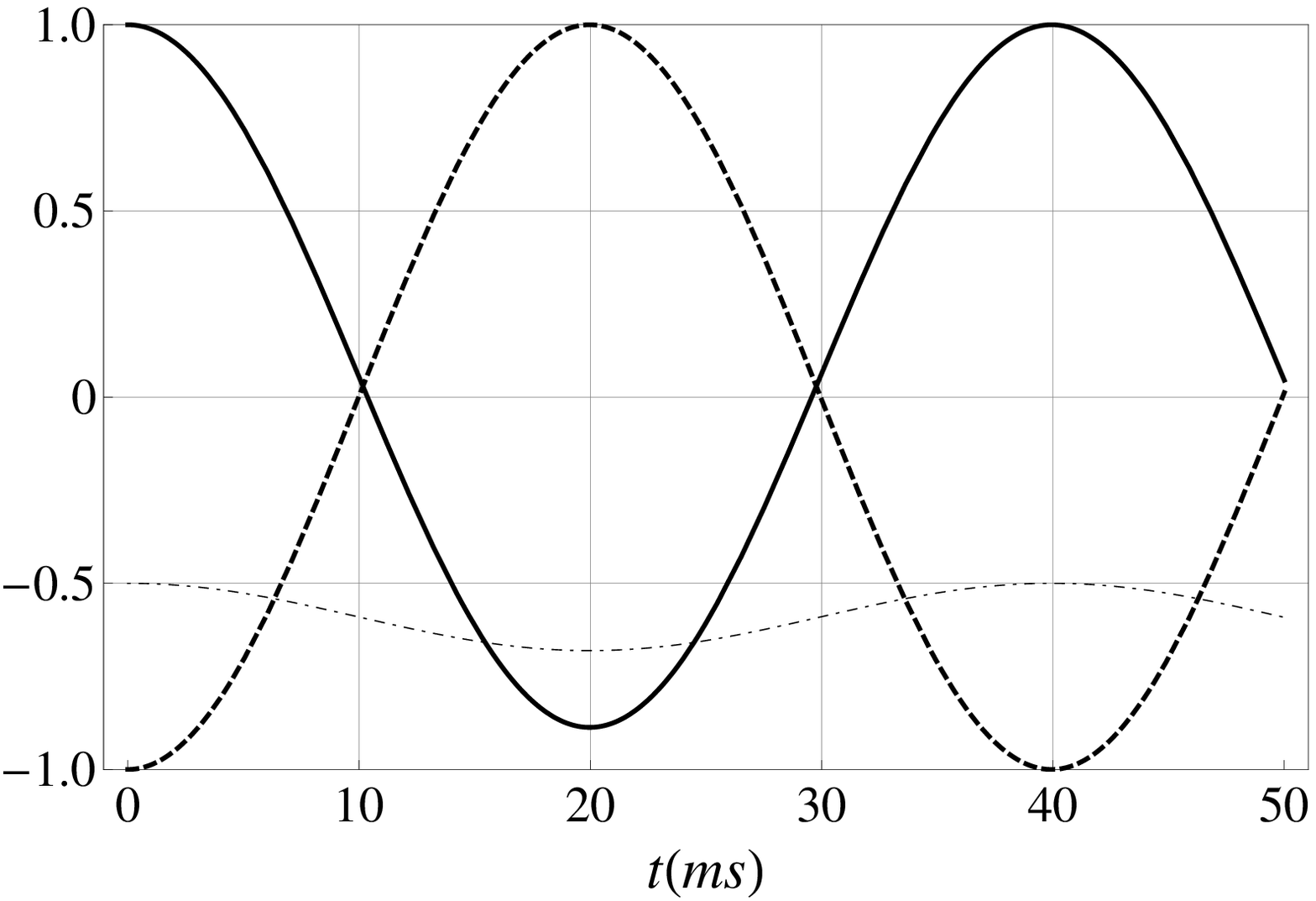} 
\caption{Current quantum dynamics of the currents for $\omega_{\Omega} = 78.3\;$rad$\cdot$Hz and $\omega_{\gamma} = 15\;$rad$\cdot$Hz. The initial condition for the current $\hat{\mathcal{I}}(t)$ (full line) is $\hat{\mathcal{I}}(0) = 1$. The initial condition for the current $\hat{\mathcal{J}}(t)$ (dashed line) is $\hat{\mathcal{J}}(0) = -1$. The initial condition for the current $\hat{\mathcal{T}}(t)$ (dot-dashed line) is $\hat{\mathcal{T}}(0) = -0.5$.}
\label{tw2}
\end{center}
\end{figure}

\begin{figure}[h]
\begin{center}
\includegraphics[scale=0.7]{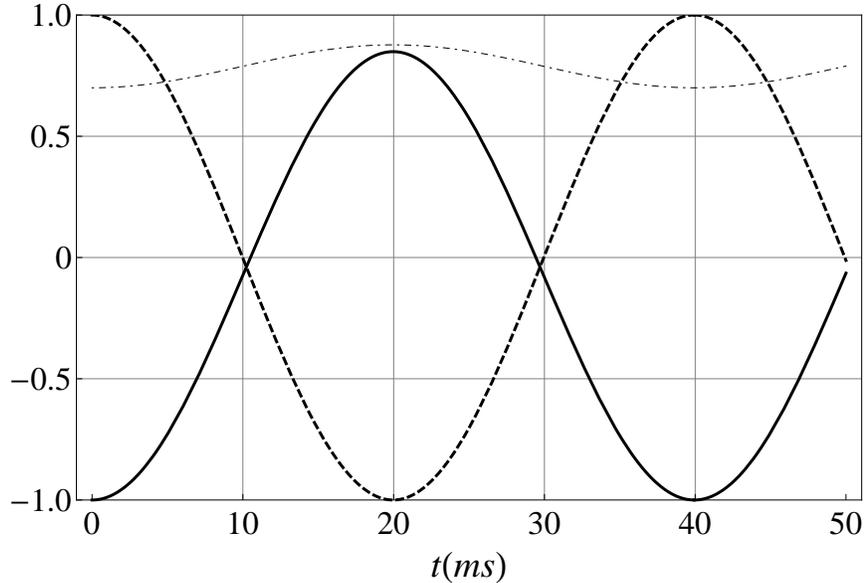}
\caption{Current quantum dynamics for $\omega_{\Omega} = 78.3\;$rad$\cdot$Hz and $\omega_{\gamma} = 15\;$rad$\cdot$Hz. The initial condition for the current $\hat{\mathcal{I}}(t)$ (full line) is $\hat{\mathcal{I}}(0) = -1.0$. The initial condition for the current $\hat{\mathcal{J}}(t)$ (dashed line) is $\hat{\mathcal{J}}(0) = 1.0$. The initial condition for the current $\hat{\mathcal{T}}(t)$ (dot-dashed line) is $\hat{\mathcal{T}}(0) = 0.7$.}
\label{tw3}
\end{center}
\end{figure}

\noindent  We get the dynamics of a simple harmonic oscillator (SHO) with natural angular frequency $\omega = \sqrt{\omega_{\gamma}^2 + 4\omega_{\Omega}^{2}}$ and period of the oscillations 
$T = \frac{2\pi}{\sqrt{\omega_{\gamma}^2 + 4\omega_{\Omega}^{2}}}$ for the current  $\hat{\mathcal{J}}$.  The Eqs. (\ref{eq1:wideeq-alf}) and (\ref{eq3:wideeq-alf}) are a system of two linear differential equations of second order. If we diagonalize the matrix  of the coefficients of the system of the Eqs. (\ref{eq1:wideeq-alf}) and (\ref{eq3:wideeq-alf}) we get the same angular frequency $\omega$. If we consider the same period of oscillation $T = 40.1\;$ms, the same angular frequency $\omega=2\pi\times 24.94\;$rad$\cdot$Hz and the same total number of particles $N = 1150$ as in the experiment  \cite{Oberthaler1},  we get the angular frequencies $\omega_{\Omega} = 78.3\;$rad$\cdot$Hz and $\omega_{\gamma} = 15\;$rad$\cdot$Hz for the parameters of the Hamiltonian. Comparing with the angular frequencies of the trap  we found $\omega_x \approx 2\pi\omega_{\Omega}$, $\omega_y \approx 2\pi \times 0.843\omega_{\Omega}$ and $\omega_z \approx 2\pi \times 1.150\omega_{\Omega}$ for the tunnelling amplitude $\Omega$ and $\omega_x \approx 2\pi\times 5.2\omega_{\gamma}$, $\omega_y \approx 2\pi \times 4.4\omega_{\gamma}$ and $\omega_z \approx 2\pi \times 6.0\omega_{\gamma}$ for the parameter $\gamma$. The height of the barrier is $V_b \approx 2\pi\hbar \times 3.36 \omega_{\Omega} \approx 2\pi\hbar \times 17.53 \omega_{\gamma}$. In the Figs. (\ref{tw2}) and (\ref{tw3}) we show the numerical solution for the same choice of these parameters. The initial condition for the first derivative for all currents is zero. The currents are normalized by $N$. For this choice of the parameters the current $\hat{\mathcal{J}}$ is independent and the initial condition determines its amplitude of oscillation. The currents $\hat{\mathcal{I}}$ and $\hat{\mathcal{T}}$ are correlated and the initial condition don't determines their amplitude of oscillation. The currents dynamics  are sensitive to the initial condition \cite{Oberthaler1} and they have the same frequency. We have self-trapping for the current $\hat{\mathcal{T}}$ and Josephson and Rabi dynamics for the currents $\hat{\mathcal{I}}$ and  $\hat{\mathcal{J}}$. 

In the limit $\alpha=\beta=\gamma=0$, the current $\hat{\mathcal{T}}$ is a conserved quantity, $[\hat{\mathcal{H}},\hat{\mathcal{T}}]=0$, but this don't means that we don't have tunnelling. We can see from Eqs. (\ref{CDN1}) and (\ref{CDN2}) that the quantum dynamics of $\hat{N}_1$, $\hat{N}_2$, and $\hat{\mathcal{I}}$ is determined by the current $\hat{\mathcal{J}}$ and the amplitude of tunnelling $\Omega$. For the currents $\hat{\mathcal{I}}$ and $\hat{\mathcal{J}}$ we get two independent SHO with $\omega_{\mathcal{I}} = \omega_{\mathcal{J}} = 2\omega_{\Omega}$  the natural angular frequency. The period of the oscillations is $T = \frac{\pi}{\omega_{\Omega}}$. In analogy with the classical SHO, the ratio between the elastic constant ${\cal K}$ and the mass $m$ is $\frac{{\cal K}}{m} = 4\omega_{\Omega}^{2}$.  The currents are  uncorrelated now and  we have Rabi oscillation for the currents $\hat{\mathcal{I}}$ and $\hat{\mathcal{J}}$. 

\section{Summary}

 I have showed that a current algebra appears when we calculate the quantum dynamics of the tunnelling of the atoms between the two condensates. I generalize the Heisenberg equation of motion to write the $n$-th time derivative of any operator. Then I calculated the quantum dynamics of these currents and I have showed that different dynamics appear when we consider different choices of the parameters of the Hamiltonian. The currents are correlated and there is interference between them and for the special choice $\alpha=\beta=\gamma=0$ they are independent. The parameters $\alpha$ and $\rho$ determines the non linearity of the interaction, the parameters $\gamma$ and $\xi$ determines the relation between the on-well energies and the external potentials, the parameter $\beta$ determines the symmetry of the interaction between the condensates.  

\section*{Acknowledgments}
The author acknowledge Capes/FAPERJ (Coordena\c{c}\~ao de Aperfei\c{c}oamento de Pessoal de N\'{\i}vel Superior/Funda\c{c}\~ao de Amparo \`a Pesquisa do Estado do Rio de Janeiro) for the financial support.



\begin{thebibliography}{10}


\bibitem{ak} J. R. Anglin and W. Ketterle, \textit{Nature} \textbf{416} (2002) 211.

\bibitem{anderson} M. H. Anderson, J. R. Ensher, M. R. Mathews, C. E. Wieman   and E. A. Cornell,  \textit{Science}  {\bf 269} (1995) 198.

\bibitem{wwcch} J. Williams, R. Walser, J. Cooper, E. A. Cornell and M. Holland, \textit{Phys. Rev. } \textbf{A 61} (2000) 033612.


\bibitem{bose} S. N. Bose, \textit{Z. Phys.}  \textbf{26} (1924) 178.

\bibitem{eins} A. Einstein, \textit{Phys. Math. K1}  \textbf{22} (1924) 261.

\bibitem{Hulet} C. A. Sackett, C. C. Bradley, M. Welling and R. G. Hulet, \text{Braz.  Jour. Phys.} \textbf{27} no. 2 (1997) 154.  

\bibitem{Dalfovo} F. Dalfovo, S. Giorgini, L. P. Pitaevskii and S. Stringari, \textit{Rev. Mod. Phys.}  \textbf{71} (1999) 463.

\bibitem{l01} A. J. Leggett, \textit{Rev. Mod. Phys.}  \textbf{73} (2001) 307.

\bibitem{Bagnato} P. W. Courteille, V. S. Bagnato and  V. I. Yukalov, \textit{Laser Phys.} \textbf{11} (2001) 659.

\bibitem{cw} E. A. Cornell and C. E. Wieman, \textit{Rev. Mod. Phys.} \textbf{74} (2002) 875.

\bibitem{Donley} E. A. Donley, N. R. Claussen, S. T. Thompson and C. E. Wieman, \textit{ Nature} \textbf{417} (2002) 529.

\bibitem{Piza} A. F. R. T. Piza, \textit{Braz. Jour. Phys.} \textbf{34} n. 3B (2004) 1102.

\bibitem{Bloch1} I. Bloch, J. Dalibard and W. Zwerger, \textit{Rev. Mod. Phys.}  \textbf{80} (2008) 875.

\bibitem{Bloch2} I. Bloch, \textit{Nature Physics} \textbf{1} (2005) 23.

\bibitem{Bloch3} M. Endres, \textit{et al}, \textit{Science} \textbf{334} (2011) 200.

\bibitem{Bloch4} I. Bloch, \textit{Nature} \textbf{453} (2008) 1016.

\bibitem{Minardi} F. S. Cataliotti, \textit{et al}, \textit{Science} \textbf{293} (2001) 843. 

\bibitem{Greene} B. D. Esry and C. H. Greene, \textit{Nature} \textbf{440} (2006) 289.

\bibitem{Kraemer} T. Kraemer, \textit{et al}, \textit{Nature} \textbf{440} (2006) 315.

\bibitem{Greiner} M. Greiner and S. F\"olling, \textit{Nature} \textbf{453} (2008) 736.

\bibitem{Oberthaler4} J. Est\`eve, \textit{et al}, \textit{Nature} \textbf{455} (2008) 1216.

\bibitem{Riedel} M. F. Riedel, \textit{et al}, \textit{Nature} \textbf{464} (2010) 1170.

\bibitem{Kinoshita} T. Kinoshita, T. Wenger and D. S. Weiss, \textit{Nature} \textbf{440} (2006) 900.

\bibitem{Thierry} T. Lahaye, \textit{et al}, \textit{Nature} \textbf{448} (2007) 672.

\bibitem{Inouye} S. Inouye, \textit{et al}, \textit{Nature} \textbf{392} (1998) 151.

\bibitem{Carusotto} I. Carusotto and C. Ciuti, \textit{Rev. Mod. Phys.}  \textbf{85} (2013) 299.


\bibitem{Ketterle1} M. R. Andrews, C. G. Townsend, H.-J. Miesner, D. S. Durfee, D. M. Kurn 
 and W. Ketterle, \textit{Science} \textbf{275} (1997) 637. 

\bibitem{Ketterle2} Y. Shin, M. Saba, T. A. Pasquini, W. Ketterle, D. E. Pritchard and A. E. Leanhardt,  \textit{Phys. Rev. Lett.} \textbf{92} (2004) 050405.

\bibitem{Oberthaler1} M. Albiez, \textit{et al}, \textit{Phys. Rev. Lett.} \textbf{95} (2005) 010402.

\bibitem{Oberthaler2} T. Zibold, \textit{et al},  \textit{Phys. Rev. Lett.} \textbf{105} (2010) 204101.

\bibitem{Oberthaler3} R. Gati, \textit{et al},  \textit{Appl. Phys.} \textbf{B 82} (2006) 207. 

\bibitem{Levy} S. Levy, E. Lahoud, I. Shomroni and J. Steinhauer, \textit{Nature} \textbf{449} (2007) 579.

\bibitem{Bloch5} S. F\"olling, \textit{et al}, \textit{Nature} \textbf{448} (2007) 1029. 

\bibitem{Josephson1} B. D. Josephson,  \textit{Phys. Lett.} \textbf{1} (1962) 251.

\bibitem{Josephson2} B. D. Josephson,  \textit{Rev. Mod. Phys.} \textbf{46} (1974) 251.

\bibitem{milb} G. J. Milburn, J. Corney, E. M. Wright and D. F. Walls, \textit{Phys. Rev. }  \textbf{A 55} (1997) 4318.

\bibitem{Javanainen} J. Javanainen, \textit{Phys. Rev. Lett.} \textbf{57} (1986) 3164.


\bibitem {l02} I. Zapata, F. Sols and A. J. Leggett, \textit{Phys. Rev.} \textbf{A 57} (1998) R28(R).

\bibitem{hines2} A. P. Hines, R. H. McKenzie and G. J. Milburn, \textit{Phys. Rev. } \textbf{A 67} (2003) 013609.

\bibitem{ours} A. P. Tonel, J. Links and A. Foerster, \textit{J. Phys. A: Math. Gen.} \textbf{38} (2005) 6879.

\bibitem{our} A. P. Tonel, J. Links and A. Foerster, \textit{J. Phys. A: Math. Gen.} \textbf{38} (2005) 1235.

\bibitem{hines} A. P. Hines, R. H. McKenzie and G. J. Milburn, \textit{Phys. Rev. } \textbf{A 71} (2005) 042303.

\bibitem{sarma} Qi Zhou and Das Sarma, \textit{Phis. Rev. } \textbf{A 82} (2010) 041601(R).

\bibitem{GSantos06a} G. Santos, A. Tonel, A. Foerster and J. Links,  \textit{Phys. Rev. } \textbf{A 73} (2006) 023609.

\bibitem{GSantos09} A. P. Tonel, C. C. N. Kuhn, G. Santos, A. Foerster, I. Roditi and Z. V. T. Santos,  \textit{Phys. Rev. } \textbf{A 79} (2009) 013624.

\bibitem{GSantos10} G. Santos, A. Foerster, J. Links, E. Mattei and S. R. Dahmen, \textit{Phys. Rev. } \textbf{A 81} (2010)  063621.

\bibitem{Gross} C. Gross, \textit{J. Phys. B: At. Mol. Opt. Phys.} \textbf{45} (2012) 103001 (20pp). 



\bibitem{GSantos11a} G. Santos, A. Foerster, I. Roditi, Z. V. T. Santos and A. P. Tonel, \textit{J. Phys. A: Math. Theor.} \textbf{41} (2008) 295003 (9pp).
 
\bibitem{GSantos11b} G. Santos, \textit{J. Phys. A: Math. Theor.} \textbf{44} (2011) 345003.

\bibitem{GSantos13} G. Santos, A. Foerster and I. Roditi, \textit{J. Phys. A: Math. Theor.} \textbf{46} (2013) 265206 (12pp).

\bibitem{jlletter} J. Links and H.-Q. Zhou, \textit{Lett. Math. Phys.} \textbf{60} (2002) 275.

\bibitem{jlreview} J. Links, H.-Q. Zhou, R. H. McKenzie and M. D. Gould, \textit{J. Phys. A: Math. Gen.} \textbf{36} (2003) R63. 

\bibitem{jlsigma} J. Links and K. E. Hibberd, \textit{SIGMA} \textbf{2} (2006) 095 (8pp).


\bibitem{Angela} A. Foerster, J. Links and H.-Q. Zhou, 
in: {\it Classical and quantum nonlinear integrable systems: theory and applications}, edited
by A. Kundu (Institute of Physics Publishing, Bristol and Philadelphia, 2003)
pp 208--233. 

\bibitem{ATonel} A. P. Tonel and L. H. Ymai, \textit{J. Phys. A: Math. Theor.} \textbf{46} (2013) 125202 (14pp).

\bibitem{GSantos06b} J. Links, A. Foerster, A. P. Tonel and G. Santos, \textit{Ann. Henri Poincar\'e}  \textbf{7}  (2006) 1591.

\bibitem{Angela2} D. Rubeni, A. Foerster, E. Mattei and I. Roditi, \textit{Nuc. Phys. } \textbf{B 856} (2012) 698.

\bibitem{GSantosBVS-arxiv} G. Santos, C. Ahn, A. Foerster and I. Roditi, \textit{Phys. Lett.} \textbf{B 746} (2015) 186.

\bibitem{jldensity} J. Links and I. Marquette, \textit{J. Phys. A: Math. Theor.} \textbf{48} (2015) 045204 (15pp).






\bibitem{Dounas} D. R. Dounas-Frazer, A. M. Hermundstad and L. D. Carr,  \textit{Phis. Rev. Lett.} \textbf{99} (2007) 200402. 

\bibitem{Graefe} Graefe, E.-M., Graney, M. and Rush, A., \textit{Phys. Rev. } \textbf{A 92} (2015) 012121.

\bibitem{GSantosCA1} G. N. Santos Filho, \textit{Current algebra for the two-site Bose-Hubbard model},  arXiv:1505.06793 [cond-mat.quant-gas].

\end{thebibliography}
\end{document}